\documentclass[12pt]{article}

\newcommand{\bbr}{I\!\! R}

\newcommand{\x}{arXiv:}
\newcommand{\m}{\mathrm}
\newcommand{\be}{\begin{equation}}
\newcommand{\ee}{\end{equation}}
\newcommand{\ba}{\begin{eqnarray}}
\newcommand{\ea}{\end{eqnarray}}

\usepackage{graphicx}
\usepackage{amssymb}
\usepackage{amsmath}
\usepackage[T1]{fontenc} %for \boldsymbol
\usepackage[ansinew]{inputenc} %for \boldsymbol
\usepackage[nosort]{cite}
\newcommand{\inbar}{\vrule height1.57ex width.4pt depth0pt}
\newcommand{\SW}{\relax{\hbox{$\ \inbar\kern-.285em{\rm S}$}}}

\begin{document}
\thispagestyle{empty}
\begin{center}

\null \vskip-1truecm \vskip2truecm

{\Large{\bf \textsf{The Special Role of Toroidal Black Holes in Holography}}}

{\large{\bf \textsf{}}}

{\large{\bf \textsf{}}}

\vskip1truecm

{\large \textsf{Brett McInnes}}

\vskip1truecm

\textsf{\\  National
  University of Singapore}

\textsf{email: matmcinn@nus.edu.sg}\\

\end{center}
\vskip1truecm \centerline{\textsf{ABSTRACT}} \baselineskip=15pt
\medskip

In the standard holographic ``dictionary'', the deep infrared of the strongly coupled boundary field theory is studied by examining the bulk region near to the event horizon of a simple AdS-Reissner-Nordstr\"{o}m black hole, near to extremality. Recently Horowitz et al. have argued that this is not correct, \emph{except} in the case of small toroidal black holes, which are therefore revealed to be particularly interesting and important. On the other hand, the Weak Gravity Conjecture postulates that black holes (including toroidal black holes) which are extremely near to extremality spontaneously emit black holes of the same kind. We show that, in the toroidal case, these ``emitted'' black holes are always small in the sense of Horowitz et al. As an application, we discuss the Grinberg-Maldacena analysis of the way one-point functions, evaluated outside an AdS-Reissner-Nordstr\"{o}m black hole, depend on the proper time of fall from the event horizon to the Cauchy horizon. We find that, for emitted toroidal black holes, this dependence effectively drops out.

\newpage

\addtocounter{section}{1}
\section* {\large{\textsf{1. Toroidal Black Holes are Special }}}
Near-extremal AdS black holes are of great interest in the AdS/CFT correspondence \cite{kn:casa,kn:nat,kn:bag}, since the region near the event horizon of such a black hole holographically represents the infra-red of the dual field theory\footnote{The near-extremal case (though \emph{not} the exactly extremal case, which we will not consider) is also distinguished as being an important stage in the evolution of many charged black holes under Hawking radiation: see \cite{kn:haoong} for a clear discussion. It may well also be of direct observational relevance, see \cite{kn:silk}.}. Such objects can be (and preferentially are) produced by a ``pair-production'' process analogous to Schwinger pair-production \cite{kn:garstrom,kn:hahoro}. This is the most straightforward way to ``construct'' such objects, and so we confine attention in this work to black holes produced in this manner.

Until very recently, it was generally accepted that the relevant bulk geometry for these black holes (in the non-rotating case) was simply that of one of the AdS$_5$-Reissner-Nordstr\"{o}m black holes\footnote{Throughout this work, for definiteness, we focus exclusively on a five-dimensional bulk.}. However, Horowitz et al. have argued strongly that, in \emph{almost} \cite{kn:horo1} all cases, this is not so: when perturbations are taken into account, the near-horizon geometries of these black holes suffer from large distortions in the near-extremal case, and the actual geometries \cite{kn:horo2} are very much more complex and difficult to handle, especially analytically.

Horowitz et al. find, however, that there is one very restricted class of exceptions: these are the ``small'' AdS$_5$ black holes with \emph{toroidal} event horizons. (``Small'' here means that the ratio $r_{\textsf{H}}/L,$ where $r_{\textsf{H}}$ gives the location of the event horizon, and $L$ is the asymptotic AdS$_5$ curvature length scale, is a small pure number\footnote{Note carefully that ``small'' does \emph{not} mean that the event horizon is small: as we will discuss in detail below, these are two separate questions.}.) In this case \emph{alone}, the standard near-extremal AdS$_5$-Reissner-Nordstr\"{o}m classical geometry \cite{kn:lemmo} is reliable. This geometry is extremely simple, particularly compared to the geometries discussed in \cite{kn:horo2}, and so (for this and other reasons \cite{kn:107}), it is of great interest: many questions can be addressed directly and exactly in this case, and only in this case.

Before we proceed to discuss this, however, we need to be a little more precise about the definition of ``smallness''. First, note that, for any given charge, the Hawking temperature of a toroidal black hole (see below) is a monotonically increasing function of $r_{\textsf{H}};$ therefore, small values of $r_{\textsf{H}}/L$ correspond to low temperatures, that is, to proximity to extremality. We will distinguish between ``QG-small'' black holes, that is, black holes with temperatures so extremely small that quantum-gravitational effects are relevant, and those which are simply ``small'', for which quantum gravity can be ignored. The toroidal black holes created by Schwinger-like black hole pair-production are QG-small; those discussed by Horowitz et al. are not, since that discussion is in the semi-classical regime.

We need this distinction because any discussion of near-extremal black holes has to take into account another phenomenon of great current interest: the \emph{Weak Gravity Conjecture} \cite{kn:motl,kn:kats,kn:NAH}) or WGC. This conjecture proposes that all black holes, \emph{including} small toroidal AdS ones, must be unstable if they are extremely close to being extremal: if, in our terminology, they are ``QG-small''. This was a surprising development, because previously quantum-gravitational effects were usually associated with extremely high-temperature objects (such as very low-mass electrically neutral, non-spinning black holes), whereas of course near-extremal black holes are very cold. It arises from the complex web of self-consistency conditions arising in string theory (see \cite{kn:swamp}), regarded as a theory of quantum gravity. That is, the WGC greatly broadens the domain in which quantum-gravity effects become important, extending it to the infra-red.

In the most interesting and consequential case \cite{kn:kats} the WGC instability causes the original black hole to emit a black hole of the same kind. Now, as mentioned, a toroidal black hole produced by a Schwinger-like mechanism will be QG-small \cite{kn:garstrom,kn:hahoro}, so we can expect it to be unstable in this way: it will emit black holes. Its temperature will either remain approximately the same\footnote{If so, $r_{\textsf{H}}$ will become \emph{smaller} as it loses charge; this can be seen easily from the formula for the Hawking temperature of these black holes, discussed below. (This is in sharp contrast with asymptotically flat black holes, which always become \emph{larger} as they lose charge, at either fixed mass or fixed temperature: see \cite{kn:109}.)} or perhaps gradually increase, with successive emissions, until it ceases to be unstable in the WGC sense. It will then merely be ``small''.

However, the status of the emitted black holes is far less clear: \emph{are they, too, necessarily small?} This is by no means obvious: at no point do we assume that the physical mass or charge of any black hole is small. Again, there is no simple relation, for toroidal black holes, between $r_{\textsf{H}}$ and the size of the event horizon. It is quite possible for the event horizon to be large (in the sense that the black hole has a large entropy) even if the ``radius'' is small.

In this note, we will show (under reasonable assumptions) that the emitted black holes in this case \emph{are in fact always small}, and so they are, or can be, exempt from the strictures of Horowitz et al. This strongly reinforces our claim that these black holes are important, and so one would wish to know more about them.

A basic question concerning these objects is the following: if indeed a given toroidal black hole is small, how can we verify this? That is, how can we demonstrate ``smallness'' by means of measurements taken in the exterior spacetime?

As we will see, it is in fact possible to compute $r_{\textsf{H}}$ if we know the size of the event horizon, together with the charge and mass of the black hole (which can be determined by observing the trajectories of particles in the exterior). However, in principle at least, there is another, more interesting, way of doing this.

It has long been hoped that it might be possible to probe the interiors of black hole event horizons, using theoretical methods (see for example \cite{kn:shenk,kn:lukasz,kn:festu}, or more recently \cite{kn:felix,kn:sam1,kn:sam2}). Ultimately even observational methods may be possible (for example, \cite{kn:ramb1,kn:ramb2}).

Recently, Grinberg and Maldacena \cite{kn:grinmald} have given a very concrete discussion of a simple question of this kind, which is directly relevant here: how is the proper time of fall from the event horizon to the Cauchy horizon\footnote{As is well known, the Cauchy horizon is in reality likely to be replaced by an actual singularity: see \cite{kn:thermong}.} of a charged (AdS) black hole reflected in quantities defined on the exterior spacetime? The suggestion in \cite{kn:grinmald} (see also \cite{kn:beren,kn:justin}) is that the large-mass behaviour of the thermal one-point function (associated with higher-derivative corrections) of a massive field outside an AdS black hole contains a term involving this time of fall quite explicitly.

It is straightforward to show (see below) that, for small toroidal charged black holes, the singularity lies ``just below the event horizon'', in the sense that the proper time of fall from the latter to the former is generically extremely short. The Grinberg-Maldacena argument gives us a way of confirming this, and therefore of confirming that a given toroidal black hole is small.

Thus, in principle at least, one can use the physics outside the black hole to confirm that it is indeed ``small'', and thus immune to the effects discussed by Horowitz et al..

We begin with a review of the geometry of charged AdS black holes, both in the general and in the less familiar toroidal cases.

\addtocounter{section}{1}
\section* {\large{\textsf{2. AdS$_5$-Reissner-Nordstr\"om Black Holes In General}}}
The principal reason for studying asymptotically AdS spacetimes is of course their role in gauge-gravity duality \cite{kn:casa,kn:nat,kn:bag}. Here, one studies a strongly-coupled field theory (which, in practice, is usually not a conformal theory) defined on a non-dynamical four-dimensional spacetime with metric $\,-\,\m{d}t^2\,+\,h^{\kappa},\,$ where $h^{\kappa}$ is a metric on a three-dimensional space of constant curvature $\kappa/X^2$; here $\kappa = 0,\,\pm 1$, and, when the space is compact (as in \cite{kn:horo1}, and as we always assume henceforth), $X$ is the characteristic length scale of the space. For example, when $\kappa = 1$ and the space is simply connected, the circumference of the three-sphere is $2\pi X$. Similarly, when $\kappa = - 1$, the injectivity radius of the relevant compact space of constant negative curvature is some multiple of $X$; and, when $\kappa = 0$ and the space is a cubic torus, then we can take $X$ to be the side length of that cube. (In all three cases, there are many possible choices of topology and geometry: see \cite{kn:conway} for the flat case. For simplicity, when $\kappa = 0,$ we assume that the space is a cubic torus.) There are AdS$_5$ black holes providing a dual bulk for all of these spacetimes \cite{kn:lemmo}; the event horizon in each case has a metric conformal to $h^{\kappa}$.

The Hawking-Page transition in the toroidal case \cite{kn:hormy,kn:surya, kn:page} (see \cite{kn:anabalon} for a recent discussion) takes an unusual form: the temperature at which it occurs is inversely proportional to $X$. Thus, if we wish to use such a black hole to explore the deep infrared of the boundary field theory, down to temperatures arbitrarily close to zero, \emph{it is essential that $X$ should be extremely large}. By this we mean that the dimensionless quantity $X/L$ must be much larger than any other dimensionless parameter in the problem, where $L$ is the asymptotic AdS curvature scale. Crucially, \emph{this includes the electromagnetic charge} on the black hole: even if the charge is large by normal standards, it should be extremely small relative to $X/L$.

To summarise: we take the non-dynamical metric $\,-\,\m{d}t^2\,+\,h^{0},\,$ and interpret $h^0$ as a metric on a three-dimensional cubic torus with a side length, $X$, which is extremely large relative to $L$. We then use this flat four-dimensional spacetime as a boundary condition for the Einstein(-Maxwell) equations in an asymptotically AdS$_5$ bulk.

Let us be more explicit. It is instructive to return temporarily to the general case. If we fix $\,-\,\m{d}t^2\,+\,h^{\kappa}\,$ (or rather the conformal structure defined by it), for any of the three possible values of $\kappa$, as a boundary condition, then \cite{kn:lemmo} a matching (AdS$_5$-Reissner-Nordstr\"{o}m) metric in the bulk, which solves the Einstein-Maxwell equations, takes the form
\begin{flalign}\label{A}
g(\m{AdSRN_5^{\kappa}})\;=\; & -\,\left(\kappa \,+\,{r^2\over L^2}\,-\,{16\pi M^*\ell_5^3\over 3r^2} +\,{4\pi k_5 Q^{*2}\ell_5^3\over 3r^4}\right)\m{d}t^2\; \notag\\  & +{\m{d}r^2\over \kappa \,+\,{r^2\over L^2}\,-\,{16\pi M^*\ell_5^3\over 3r^2}\,+\,{4\pi k_5 Q^{*2}\ell_5^3\over 3r^4}} \,+\,{r^2\over L^2}\,h^{\kappa}.
\end{flalign}
Here $L$ is the AdS$_5$ scale as above, $M^*$ and $Q^*$ are mass and charge parameters (proportional but not equal to the physical mass $M$ and the physical charge $Q$, see below), $k_5$ is the five-dimensional Coulomb constant (with units of length, unlike its four-dimensional counterpart), and $\ell_5$ is the gravitational length scale\footnote{We use units in which $M^*$ has units of inverse length, $Q^*$ is dimensionless and so is entropy. We never use Planck units.} for AdS$_5$.

However, this metric only solves the equations in the $\kappa = \pm 1$ cases if the spatial length scale $X$ in $h^{\kappa}$ is precisely equal to $L$. That is: in these cases, \emph{the boundary length scale is forced to coincide with the bulk curvature scale}.

In sharp contrast, when $\kappa = 0$, there is no such requirement, so $X$ is in principle entirely independent of $L$ (though, as we saw, there are physical reasons for requiring that $X$ must be large relative to $L$). This means, in effect, that there is a \emph{new independent parameter} in the problem in the $\kappa = 0$ case, and we have to deal with this.

With this point understood, from this point onwards we only consider the $\kappa = 0$ case in equation (\ref{A}); that is, we consider only metrics of the form
\begin{flalign}\label{AA}
g(\m{AdSRN_5^{0}})\;=\; & -\,\left({r^2\over L^2}\,-\,{16\pi M^*\ell_5^3\over 3r^2} +\,{4\pi k_5 Q^{*2}\ell_5^3\over 3r^4}\right)\m{d}t^2\; \notag\\  & +{\m{d}r^2\over {r^2\over L^2}\,-\,{16\pi M^*\ell_5^3\over 3r^2}\,+\,{4\pi k_5 Q^{*2}\ell_5^3\over 3r^4}} \,+\,{r^2\over L^2}\,h^{0}.
\end{flalign}
We note in passing that the Hawking temperature in this case is given by
\begin{equation}\label{AAA}
T^0\;=\;{1\over \pi}\left({r_{\textsf{H}}\over L^2}\;-\;{2\pi k_5Q^{* 2}\ell_5^3 \over 3 r_{\textsf{H}}^5}\right),
\end{equation}
which, as mentioned earlier, is a monotonically increasing function of $r_{\textsf{H}}$, and therefore of $M^*$, for given $Q^*.$ Thus, these black holes always have a positive specific heat \cite{kn:ruong}.

We now briefly review the unusual status of Cosmic Censorship and of the WGC for these black holes.

\addtocounter{section}{1}
\section* {\large{\textsf{3. Censorship and the Weak Gravity Conjecture for Toroidal Black Holes}}}
We now wish to consider the Weak Gravity Conjecture \cite{kn:motl,kn:kats,kn:NAH}, which is closely related to the idea that very near-extremal black holes must decay, either by emitting particles/branes, or by emitting another black hole. The latter process, which is the one that concerns us here, is associated \cite{kn:107}, at least at the classical level, with the production of a naked singularity; that is, with a violation of Cosmic Censorship.

It is important to stress the remarkable power of the theorem (see \cite{kn:wald}, Theorem 12.2.1) used in \cite{kn:107} to deduce the presence of a naked singularity here. This theorem does not require any assumptions as to how the new black hole comes into existence: its assumptions only concern some very general aspects of spacetime causal structure (such as the connectedness of the interiors of lightcones). Physically, there is a distinction between a black hole created by a semi-classical splitting of an event horizon, and one which emerges from a quantum fluctuation of some kind; but the theorem is equally applicable to both cases. (The theorem likewise requires no energy condition to be satisfied.) Thus, the existence of the naked singularity is assured despite the uncertainty as to the precise details of how the new black hole comes into existence.

The status of Censorship, both in asymptotically flat \cite{kn:ulrich} and in asymptotically AdS \cite{kn:toby,kn:weak} spacetimes, has been much debated recently: see \cite{kn:ongcensor} for an overview. Here we follow the point of view advocated in \cite{kn:roberto} (see also \cite{kn:newemp}): naked singularities probably are produced in certain highly dynamical spacetimes, but they do not persist, and so Censorship is a reliable guide to the structure of the \emph{final} state. That is, we will assume that the emitted black hole is indeed a black hole and not a naked singularity.

Censorship in this case implies the existence of horizons of the black hole with metric given in (\ref{AA}). The outer horizon, located at $r = r_{\textsf{H}},$ will be a flat cubic torus of side length $\Delta,$ where, from equation (\ref{A}), $\Delta = r_{\textsf{H}}X/L,$ and where $X$ was defined in the preceding Section.

This event horizon ``radius'' is related to the mass and charge of the black hole in a way which may be unfamiliar, so let us review it.

For later convenience we define a dimensionless quantity $W$ by $W \equiv (X/L)^3;$ according to our discussion above, $W$ is to be regarded as an extremely large (but finite) dimensionless parameter. The side length of the horizon is then
\begin{equation}\label{B}
\Delta \;=\; W^{1/3}r_{\textsf{H}}.
\end{equation}
It must be stressed that $\Delta$ should be regarded as an \emph{observational} parameter: an observer in the spacetime who is willing to venture sufficiently near to the event horizon can measure its side length \emph{directly}. (In addition, the entropy of the black hole is proportional to $\Delta^3$, again underlining the ``physicality'' of this quantity.) One should think of it as a new physical parameter describing the black hole, like (but independent of) the mass and the charge. We underline once more that, even if $\Delta$ is large, it by no means follows that the same is true of $r_{\textsf{H}};$ in fact, it turns out (if we fix the black hole mass and charge) that $r_{\textsf{H}}$ is small either when $\Delta$ is small or when it is large.

In fact, a measurement of $\Delta$ will be essential in order to determine $r_{\textsf{H}},$ since, in contrast to the $\kappa = \pm 1$ cases, the latter is \emph{not} fixed even if the physical mass $M$ and physical charge $Q$ of the black hole are known\footnote{If we do not compactify the three-dimensional space, taking it to have the topology of flat simply-connected $\bbr^3,$ then the situation is radically different. The metric (\ref{AA}) is of course still a valid solution of the Einstein equation, but now we have nothing analogous to $\Delta, M,$ and $Q$ (all of which are now formally infinite) to help us to fix $M^*$ and $Q^*$. In this case, one has to rely on holography itself to do this, by relating these parameters to well-defined physical parameters (chemical potential, enthalpy density and so on) describing the boundary field theory.}. Instead, $r_{\textsf{H}}$ is a definite function of $M, Q, $ \emph{and} $\Delta,$ given in the following manner.

First, it turns out (see \cite{kn:106} for an elementary derivation) that the parameters $M^*$ and $Q^*$ are related to the physical mass and charge through $M^* = M/W,$ and $Q^* = Q/W$. We therefore have
\begin{equation}\label{C}
{r_{\textsf{H}}^2\over L^2}\,-\,{16\pi M\ell_5^3\over 3Wr_{\textsf{H}}^2}\,+\,{4\pi k_5 Q^2\ell_5^3\over 3W^2r_{\textsf{H}}^4} \;=\; 0.
\end{equation}
We can then solve for $r_{\textsf{H}}$ by eliminating $W$ between equations (\ref{B}) and (\ref{C}). We will return to this, below.

Notice from this discussion that, by taking $W$ sufficiently large ---$\,$ recall that we do wish to do this, to describe the deep infrared of the boundary field theory ---$\,$ we can always force $M^*$ and $Q^*$ to be very small, for any toroidal black hole with given mass and charge. Even if the actual physical charge is large, then, \emph{we should think of $Q^*$, in particular, as a very small dimensionless parameter.}

Black holes with toroidal event horizons do not exist in the asymptotically flat case, and so they are not well-behaved in the limit $L \rightarrow \infty.$ They form a disjoint branch of solutions, and can therefore behave in unexpected ways. An important example \cite{kn:106} is the explicit condition for (classical) Cosmic Censorship to hold here:
\begin{equation}\label{D}
{M\ell_5\over Q}\;=\;{M^*\ell_5\over Q^*}\;\geq \;{3\over 16}\left({12\,k_5^2\over \pi L^2}\right)^{{1\over 3}}\,Q^{*{1\over 3}}.
\end{equation}
This is very different from the asymptotically flat case, where of course the mass to charge ratio is bounded below by a fixed constant if Cosmic Censorship is to hold. Notice in particular that, if $Q^*$ is small (which in fact it normally is here, as we have stressed), then the dimensionless quantity $M\ell_5/Q$ can be small \emph{without} violating classical Censorship, something which is not possible in the asymptotically flat case.

The Weak Gravity Conjecture also takes an unusual form here \cite{kn:106}. Let us suppose that we have a very near-extremal toroidal black hole, with mass $M$ and charge $Q$. We assume, in accordance with the WGC \cite{kn:motl,kn:kats,kn:NAH}, that it emits a toroidal black hole with smaller parameters $m < M$ and $q < Q.$ (We assume throughout that neither $Q$ nor $q$ vanishes.) Then it is possible to show \cite{kn:106} that, if the original black hole continues to satisfy Censorship after this bifurcation, we must have
\begin{equation}\label{E}
{m\ell_5\over q}\;=\;{m^*\ell_5\over q^*}\; < \;{1\over 4}\left({12\,k_5^2\over \pi L^2}\right)^{{1\over 3}}\,Q^{*{1\over 3}}.
\end{equation}

As is well known, the analogue of this relation in the asymptotically flat case \cite{kn:kats} is simply the statement that classical Censorship must be violated by the emitted black hole, which motivates the claim that the bifurcation can only be understood with the aid of a \emph{quantum-gravitational} modification of Censorship itself. (This is why the WGC instability only affects ``QG-small'' black holes.) But it was shown in \cite{kn:107} \emph{that this is not true in the toroidal case}: the replacement of the $3/16$ factor on the right side of (\ref{D}) by the $1/4$ factor in (\ref{E}) means that it is possible (though not compulsory) for the emitted black hole to satisfy classical Censorship\footnote{More concretely: because $Q^*$ is extremely small, the inequality (\ref{E}) means that $m\ell_5$ is also very small relative to $q$. However, as we stressed earlier, in the toroidal case this does not necessarily violate classical Censorship.}. Because it can be analysed explicitly, that is, without understanding the details of quantum gravity, this is the case on which we focus here.

We are now in a position to establish that, when a toroidal black hole is emitted in accordance with the WGC, it must be small.

\addtocounter{section}{1}
\section* {\large{\textsf{4. Emitted Toroidal Black Holes are Always Small}}}
As mentioned earlier, equations (\ref{B}) and (\ref{C}) can be combined to eliminate $W$, and the result (for the emitted black hole, indicated by the superscript) is
\begin{equation}\label{F}
r^{\textsf{E}}_{\textsf{H}}\left(\Delta^{\textsf{E}}\right)\;=\;{16\pi m\ell_5^3 \over 3\left({\left(\Delta^{\textsf{E}}\right)^3\over L^2}\;+\;{4\pi k_5 q^2\ell_5^3\over 3\left(\Delta^{\textsf{E}}\right)^3}\right)}.
\end{equation}
For fixed $m$ and $q$, then, $r^{\textsf{E}}_{\textsf{H}}$ is a simple function of $\Delta^{\textsf{E}}.$ Assuming that $q \neq 0,$ one finds that this function is bounded above; it is small for small $\Delta^{\textsf{E}},$ but \emph{also} for large $\Delta^{\textsf{E}}.$

An elementary computation shows that, for all $\Delta^{\textsf{E}},$
\begin{equation}\label{G}
{r^{\textsf{E}}_{\textsf{H}}\over L}\;\leq \; {4\sqrt{\pi}m\ell_5^{3/2}\over \sqrt{3k_5}q}.
\end{equation}

Combining this with the inequality (\ref{E}), we now have
\begin{equation}\label{H}
{r^{\textsf{E}}_{\textsf{H}}\over L}\;\leq \; \left({16\pi k_5\over 3L}\right)^{{1\over 6}}\,Q^{*{1\over 3}},
\end{equation}
where, as above, $Q^*$ is the charge parameter of the original toroidal black hole.

Since, for reasons we have explained, $Q^*$ is a small pure number, it follows that \emph{the emitted black hole must be small} in the sense of Horowitz et al.; that is, it accurately represents the deep infrared of the dual field theory. According to \cite{kn:horo1,kn:horo2}, these emitted black holes are by far the simplest black holes of which this can be said, and so their properties are of considerable interest. Notice that the bulk black hole is ``small'' as a result, paradoxically, of the fact that the boundary torus is very \emph{large}.

Recall that we have not assumed that either $Q$, the actual physical charge of the original black hole, or $q$, that of the emitted one, are small by normal standards. They could easily be very large (which is why the smallness of $r^{\textsf{E}}_{\textsf{H}}$ is non-trivial). This raises the question, however, as to whether the gravitational field in the vicinity of the event horizon of the emitted black hole is necessarily large if the charge is large.

The answer is: not necessarily. Suppose that we fix the mass and charge of the emitted black hole, and assume a large value of $\Delta^{\textsf{E}}$; as explained above, this is compatible with a small value for $r^{\textsf{E}}_{\textsf{H}}$. (One might argue in fact that this larger value for $\Delta^{\textsf{E}}$ should be favoured thermodynamically, since of course it is the value that leads to a (much) larger entropy for the emitted black hole). The Kretschmann invariant at the event horizon, $\textsf{Kr}(r^{\textsf{E}}_{\textsf{H}})$ (measured relative to the asymptotic AdS$_5$ Kretschmann invariant $\textsf{Kr}(AdS_5)$), is
\begin{equation}\label{K}
{\textsf{Kr}(r^{\textsf{E}}_{\textsf{H}})\over \textsf{Kr}(AdS_5)}\;=\;1\;+\;{4\pi k_5\ell_5^3L^2Q^2\over 15\left(\Delta^{\textsf{E}}\right)^6} \;+\;\mathcal{O}\left(\Delta^{\textsf{E}}\right)^{-8},
\end{equation}
so the geometry at the event horizon differs little from the asymptotic AdS$_5$ geometry, even if $r^{\textsf{E}}_{\textsf{H}}$ happens to be small, as it is here. The fact that the ``radius'' is small does \emph{not} mean that the curvature is large just outside the event horizon.

We now briefly discuss one particular application of these results.

\addtocounter{section}{1}
\section* {\large{\textsf{5. Just Under the Event Horizon}}}
It is well known that exact Reissner-Nordstr\"{o}m spacetimes have \emph{Cauchy horizons} in their interiors. The fate of these objects when perturbations, both classical and quantum-mechanical, are taken into account, remains a matter of very current debate: see for example \cite{kn:boy1,kn:boy2}, and references therein. The simplest possibility is that the Cauchy horizon becomes singular in some way \cite{kn:daf}. In order to make contact with the work of Grinberg and Maldacena \cite{kn:grinmald}, we will work with this assumption: that is, we interpret the ``proper time of fall to the singularity'' to mean the time to fall from the outer horizon to the idealised location of the Cauchy horizon.

Marolf \cite{kn:marolf} has pointed out that, in the case of near-extremal black holes, a singularity located at the erstwhile Cauchy horizon is extremely close to the outer horizon, in the sense that an object (with an energy per unit mass which is not very small) falling through the event horizon will reach the singularity almost instantly. (See \cite{kn:sofia} for a recent discussion of this and related claims.) This is interesting for a variety of reasons. One such reason is that the spacetime curvature at and just outside the outer horizon can be very small; so ``closeness'' to a spacetime singularity, even if it be a strong curvature singularity, need \emph{not} imply the presence of intense gravitational fields. Quantum effects of the singularity may nevertheless be detectable outside the horizon, and of course this would be of the utmost interest. Let us discuss these questions in the case of small toroidal AdS$_5$-Reissner-Nordstr\"{o}m black holes.

Following \cite{kn:marolf}, we consider a particle of non-zero energy per unit mass $\gamma$ falling into the emitted black hole. Here we have to consider the simple fact that the ``proper time of fall from the event horizon to the Cauchy horizon'' is of course not uniquely defined: it depends on $\gamma$. However, we can fix $\gamma$ at some typical value (bearing in mind that, generically, the energy per unit mass of particles falling across event horizons is not small), and study the time of fall for the specific black hole in which we are interested here, as compared with other metrics in some general class.

In order to make such a comparison, let us consider any charged black hole metric which is such that, in Reissner-Nordstr\"{o}m-like coordinates, the $tt$ and $rr$ components satisfy $g_{tt}g_{rr} = -1$ (see \cite{kn:ted}); obviously the metrics in (\ref{AA}) are of this kind. Then the trajectory of a (not necessarily free) particle with energy per unit mass $\gamma$ falling through the region between the event and Cauchy horizons satisfies $\left({\m{d}r\over \m{d}\tau}\right)^2 = g_{tt} + \gamma^2,$ where $\tau$ is proper time along the worldline and where $g_{tt} \geq 0$ in that region. Consequently, if $r^{\textsf{E}}_{\textsf{C}}$ denotes the location of the Cauchy horizon, then the proper time of fall $\tau$ satisfies (in the case of free fall, in which $\gamma$ is constant)
\begin{equation}\label{I}
\tau \;\leq\;{r^{\textsf{E}}_{\textsf{H}} - r^{\textsf{E}}_{\textsf{C}}\over \gamma},
\end{equation}
and so $\tau < r^{\textsf{E}}_{\textsf{H}}/\gamma,$ whatever the value of $r^{\textsf{E}}_{\textsf{C}}$ may be. This is useful, because perturbations will convert the Cauchy horizon to a singularity and may well cause it to shift position (see for example \cite{kn:boy1,kn:boy2}). Thus finally we have, from the inequality (\ref{H}),
\begin{equation}\label{J}
{\tau \over L}\; < \; {1\over \gamma}\left({16\pi k_5\over 3L}\right)^{{1\over 6}}\,Q^{*{1\over 3}}.
\end{equation}
Again, the extreme smallness of $Q^*$ means that $\tau$ is very small relative to $L$.

There is no reason to expect the ``radius'' of the event horizon of a general black hole to be particularly small, and so the proper time of free fall can be large even if $\gamma$ is large. But in the case of the black holes we are studying here, we know that $r^{\textsf{E}}_{\textsf{H}}$ is indeed small; so the proper time of free fall from event horizon to Cauchy horizon (or whatever replaces it in the perturbed case) must likewise be small, unless $\gamma$ is fine-tuned to be very small. (That is, for a given value of $\gamma,$ $\tau$ is unusually small for this variety of black hole, compared to other black holes.)

In short: for (classical) toroidal black holes emitted by ``primordial'' toroidal black holes in accordance with the WGC, \emph{the singularity lies ``just under'' the event horizon}: a generic object falling through the event horizon would meet the singularity almost instantly.

This has interesting consequences for the work of Grinberg and Maldacena \cite{kn:grinmald}. In their discussion\footnote{Only spherical black holes are considered in Section 6 of \cite{kn:grinmald}. However, this does not appear to affect the specific point we are raising here.} of the large-mass behaviour of the thermal one-point function of a massive field outside an AdS$_5$-Reissner-Nordstr\"{o}m black hole (Section 6 of \cite{kn:grinmald}), they find that the one-point function continues to depend on the time of fall, no matter how close the black hole is to being extremal. At first sight this is surprising, because the ``neck'' region characteristic of these black holes becomes elongated in the very near-extremal limit.

In the toroidal case, if we wish to go extremely close to extremality without encountering a Hawking-Page transition, then we need to take $W$ to be extremely large. This will reduce $Q^*$ to negligible proportions, and then the inequality (\ref{J}) means that the time of fall essentially drops out of the formula for the one-point function. This seems to resolve the puzzle in this particular case.

\addtocounter{section}{1}
\section* {\large{\textsf{6. Conclusion}}}
The results of \cite{kn:horo1,kn:horo2} are of fundamental importance to the application of holographic techniques to the deep infrared of strongly coupled field theories. They present us with a choice:

$\bullet$ Study the field theory on a flat torus (which, fortunately, can and in fact should be taken to be extremely large, so large indeed that it is essentially ordinary flat space); then one can use the ordinary AdS$_5$ toroidal Reissner-Nordstr\"{o}m black hole geometry for the bulk, given in equation (\ref{AA}) above, provided that it is sufficiently ``small''; \emph{or}

$\bullet$ Study the field theory on some other spatial geometry, but then be prepared to use a very intricate bulk spacetime geometry, as described in detail in \cite{kn:horo2}.

Our results in this work show that the first option arises extremely naturally in the context of the Weak Gravity Conjecture: the postulated decay of toroidal black holes in that scenario naturally (in fact, necessarily) gives rise to ``small'' toroidal black holes. One might say that the small toroidal black holes needed for holographic purposes are generically those found ``in nature'' in the AdS context.

\addtocounter{section}{1}
\section*{\large{\textsf{Acknowledgements}}}
The author is grateful to Prof. Ong Yen Chin for a very helpful discussion, and to Dr Soon Wanmei for inspiration.

\end{document}